\documentclass[prd,twocolumn,superscriptaddress,altaffilletter,showpacs,nofootinbib]{revtex4}
\usepackage[dvips]{graphicx}
\usepackage{amsmath}

\newcommand{\be}{\begin{equation}}
\newcommand{\ee}{\end{equation}}
\newcommand{\bea}{\begin{eqnarray}}
\newcommand{\eea}{\end{eqnarray}}

\begin{document}

\title{Comment on ``Extended Born-Infeld theory and the bouncing magnetic universe''}

\author{Ricardo Garc\'{\i}a-Salcedo}\email{rigarcias@ipn.mx}\affiliation{Centro de Investigacion en Ciencia Aplicada y Tecnologia Avanzada - Legaria del IPN, M\'exico D.F., CP 06050, M\'exico.}

\author{Tame Gonzalez}\email{tamegc72@gmail.com}\affiliation{Departamento de Ingenier\'ia Civil, Divisi\'on de Ingenier\'ia, Universidad de Guanajuato, Guanajuato, CP 36000, M\'exico.}

\author{Israel Quiros}\email{iquiros6403@gmail.com}\affiliation{Departamento de Matem\'aticas, Centro Universitario de Ciencias Ex\'actas e Ingenier\'ias, Universidad de Guadalajara, Guadalajara, CP 44430, Jalisco, M\'exico.}

\date{\today}

\begin{abstract}
In a recent paper [Phys. Rev. D{\bf 85}, 023528 (2012)] the authors proposed a generalized Born-Infeld electrodynamics coupled to general relativity which produces a nonsingular bouncing universe. For a magnetic universe the resulting cosmic evolution inevitably interpolates between asymptotic de Sitter states. Here we shall show that (i) the above theory does not have the standard weak field Maxwell limit, (ii) a sudden curvature singularity -- not better than the big bang -- arises, (iii) the speed of sound squared is a negative quantity signaling instability against small perturbations of the background energy density, and that (iv) the conclusion about the inevitability of the asymptotic vacuum regime in a magnetic universe is wrong.
\end{abstract}

\pacs{98.80.Cq}
\maketitle

\section{Introduction}

Studying the equations of the non-linear electrodynamics (NLED) is an attractive subject of research in general relativity thanks to the fact that such quantum phenomena as vacuum polarization can be implemented in a classical model through their impact on the properties of the background space-time. Even if the NLED models coupled to general relativistic cosmology describe hypothetical systems reminiscent of the fields in the real world, these models comprise interesting dynamical behavior that is worthy of independent investigation. 

The prototype of a NLED theory is provided by the original Born-Infeld Lagrangian \cite{born-infeld}: 

\bea L=-\gamma^2\left(\sqrt{1+F/2\gamma^2}-1\right),\label{l-b-i}\eea where $F$ stands for the electromagnetic (EM) invariant $F:=F^{\mu \nu}F_{\mu \nu }$, and $\gamma$ is a free constant parameter. The motivation of the authors was to have regular field configurations without singularities. The gravitational field was not included in their analysis. If one introduces gravitational effects through the theory of general relativity, a drawback of Born-Infeld proposal unfolds: there is no place for a regular cosmological scenario with the combined effects of gravity and NLED. 

In the reference \cite{novello-prd-2012}, motivated by the original Born-Infeld's idea of having regular field configurations with the EM field bounded -- this time in a magnetic universe -- the authors focused in a modification of the Lagrangian (\ref{l-b-i}) by the inclusion of a term quadratic in the field $F$ within the square root. Besides, in \cite{novello-prd-2012} the Lagrangian term $\propto \gamma^2$ in (\ref{l-b-i}) was removed. According to the authors, the resulting cosmological model is characterized by a bounce at some non vanishing value of the scale factor and, the corresponding cosmological evolution inevitably interpolates between asymptotic vacuum (de Sitter) states.

In this comment we shall show that the latter conclusion is wrong and -- what is more relevant -- that their theory does not have the standard weak field (linear) Maxwell limit. Besides, a sudden curvature singularity is inevitable in the resulting cosmological model which, together with the fact that the speed of sound squared can be a negative quantity -- signaling instability against small perturbations of the magnetic background -- leads to concluding that the theory should be ruled out.

\section{Nonlinear Electrodynamics coupled to General Relativity}

The Einstein-Hilbert action of gravity coupled to NLED is given by

\be S=\int d^4x\sqrt{-g}\left[ R+L_{\text{m}}+L(F,G)\right],\label{action}\ee where $R$ is the curvature scalar, $L_{\text{m}}$ -- the Lagrangian of the background matter, and $L(F,G)$ is the gauge-invariant electromagnetic (EM) Lagrangian, which is a function of the EM invariants $F:=F^{\mu \nu}F_{\mu \nu }$ and $G:=\frac{1}{2}\epsilon_{\alpha\beta\mu\nu}F^{\alpha\beta}F_{\mu\nu}$ (see, for instance, Ref. \cite{klippert-prd-2002}). As usual, the EM tensor is defined as $F_{\mu\nu}:=A_{\nu,\mu}-A_{\mu,\nu}$ (here the comma denotes partial derivative in respect to the spacetime coordinates, while the semicolon denotes covariant derivative instead). Standard (linear) Maxwell electrodynamics is given by the Lagrangian $L(F)=-F/4$. 

The gravitational field equations can be derived from the action (\ref{action}) by performing variations with respect to the spacetime metric $g_{\mu\nu}$, to obtain: $G_{\mu\nu}=T_{\mu\nu}^{\text{m}}+T_{\mu\nu}^{\textsc{em}}$, where $T_{\mu\nu}^{\text{m}}=\left(\rho_{\text{m}}+p_{\text{m}}\right) u_{\mu}u_{\nu}-p_{\text{m}}g_{\mu\nu}$, and $T_{\mu\nu}^{\textsc{em}}=g_{\mu\nu}\,\left[L(F)-G L_G\right]-4F_{\mu\alpha}F_\nu^{\;\;\alpha}\,L_F$, with $\rho_{\text{m}}=\rho_{\text{m}}(t)$, $p_{\text{m}}=p_{\text{m}}(t)$ -- the energy density and barotropic pressure of the background fluid, respectively, while $L_F\equiv dL/dF$, $L_{FF}\equiv d^2L/dF^2$, etc. Variation with respect to the components of the EM potential $A_\mu$ yields to the EM field equations $(F^{\mu\nu}\,L_F+\epsilon^{\alpha\beta\mu\nu}F_{\alpha\beta}L_G/2)_{;\mu}=0$.

In this comment we shall consider a homogeneous and isotropic Friedmann-Robertson-Walker (FRW) background metric with flat spatial sections: $ds^{2}=-dt^{2}+a(t)^{2}\delta_{ij}dx^idx^j$ ($i,j=1,2,3$), where $a(t)$ is the cosmological scale factor. In order to meet the requirements of homogeneous and isotropic cosmology, the energy density and the pressure of the NLED field should be evaluated by averaging over volume. To do this, we define the volumetric spatial average of a quantity $X$ at the time $t$ by  (for details see, for instance, \cite{tolman, gasperini}): $$\overline{X}\equiv \lim_{V\rightarrow V_{0}}\frac{1}{V}\int d^3x \sqrt{-g}\;X,$$ where $V=\int d^3x\sqrt{-g}$ and $V_{0}$ is a sufficiently large time-dependent three-volume. Following the above averaging procedure, for the electromagnetic field to act as a source for the FRW model we need to impose that (the Latin indexes run over three-space); 

\bea &&\overline{E}_{i}=0,\;\overline{B}_{i}=0,\;\overline{E_{i}B_{j}}=0,\nonumber\\
&&\overline{E_{i}E_{j}}=-\frac{1}{3}E^{2}g_{ij},\;\overline{B_{i}B_{j}}=-\frac{1}{3}B^{2}g_{ij}.\nonumber\eea Additionally it has to be assumed that the EM fields, being random fields, have coherent lengths that are much shorter than the cosmological horizon scales. Under these assumptions the energy-momentum tensor of the EM field can be written in the form \cite{klippert-prd-2002}: $T_{\mu\nu}^{\textsc{em}}=\left(\rho_{\textsc{em}}+p_{\textsc{em}}\right) u_{\mu}u_{\nu}-p_{\textsc{em}} g_{\mu\nu}$, where $\rho_{\textsc{em}}=-L+G L_G-4L_F E^{2}$, and $p_{\textsc{em}}=L-G L_G-4(2B^{2}-E^{2}) L_F/3$, with $E$ and $B$ being the averaged electric and magnetic fields, respectively.


\begin{figure*}[t!]\begin{center}
\includegraphics[width=5.5cm,height=5cm]{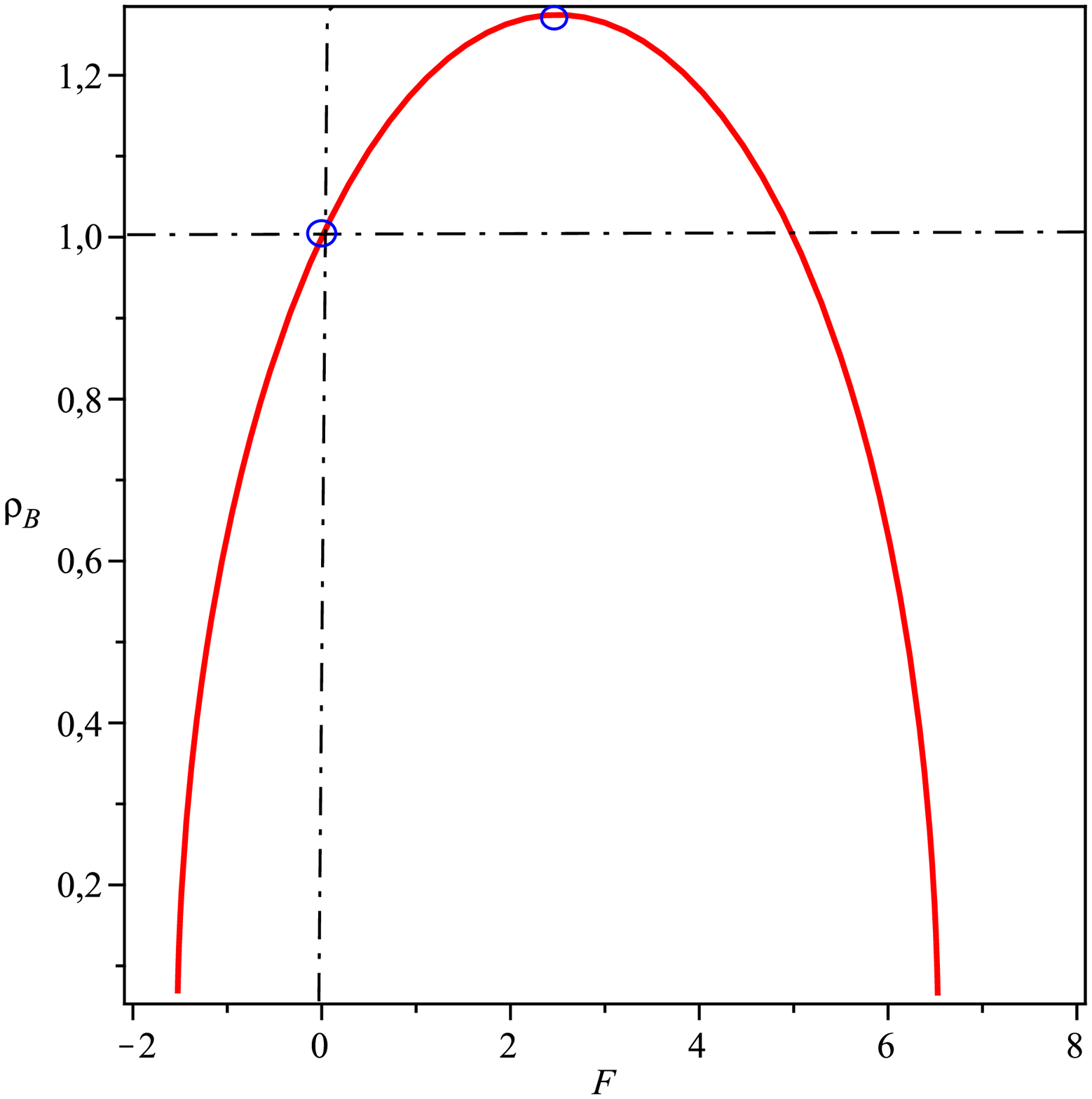}
\includegraphics[width=5.5cm,height=5cm]{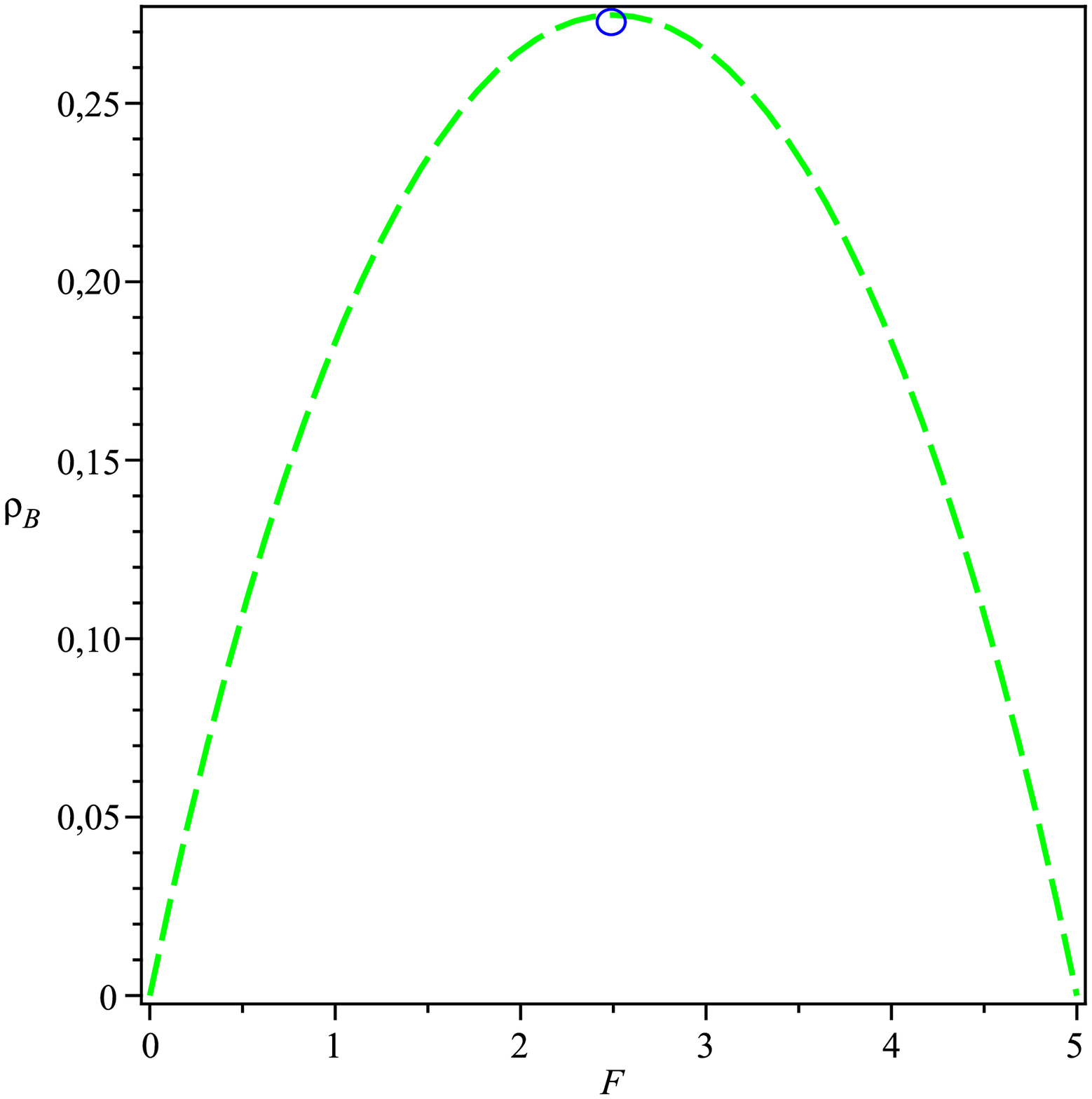}
\includegraphics[width=5.5cm,height=5cm]{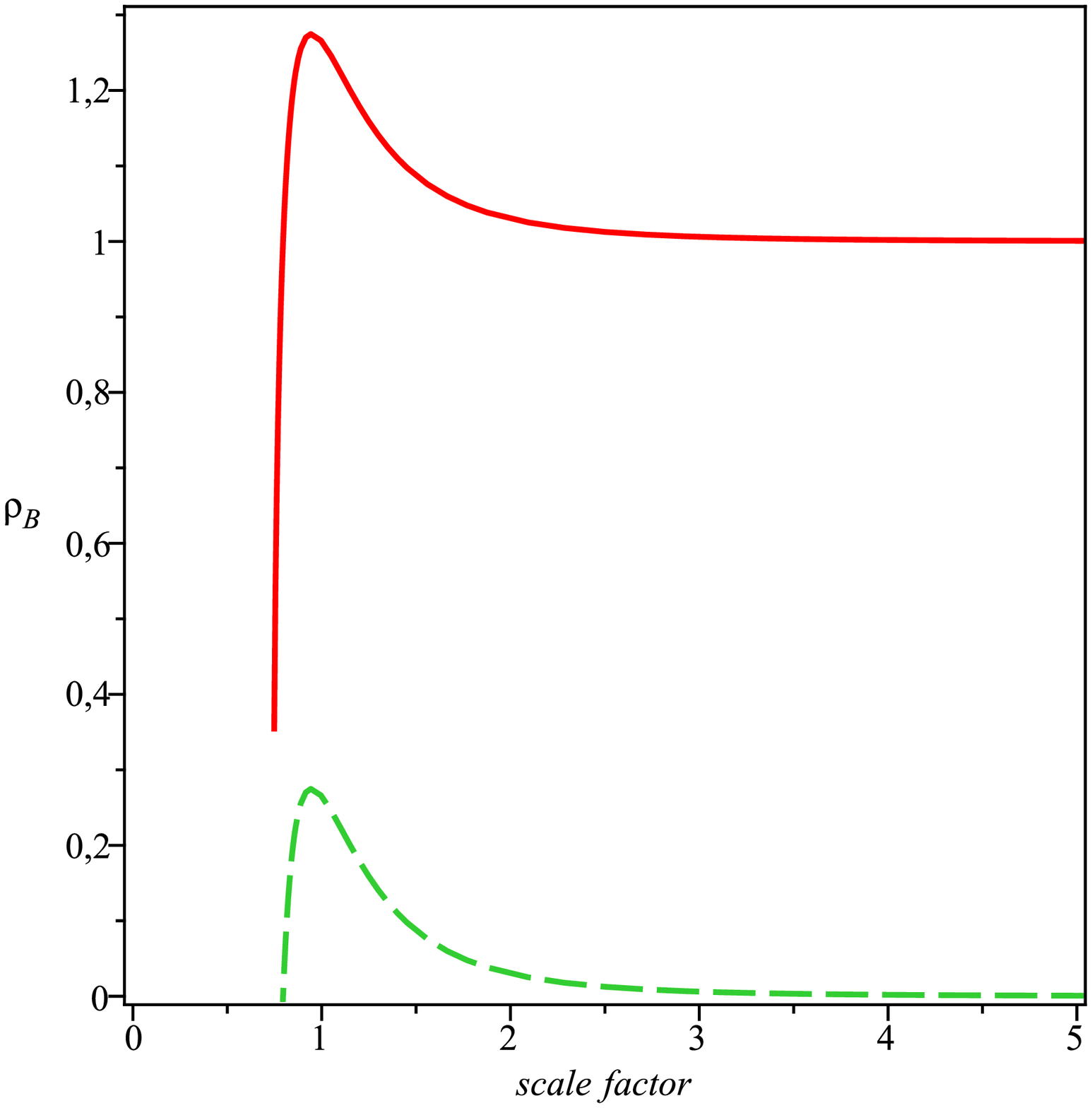}
\end{center}\vspace{0.3in}
\caption{Plot of the energy density of the magnetic field $\rho_\textsc{b}$ vs $F$ for arbitrarily chosen $\alpha^2=0.1$ and $\gamma^2=1$ (left-hand and center figures), and of $\rho_\textsc{b}$ vs scale factor $a$ (right-hand figure). The solid curve is for the theory of Ref. \cite{novello-prd-2012}, which is given by equations (\ref{l-model}), and (\ref{rho-vs-f}), while the dashed curve is for the model (\ref{l-model'}). For the theory of \cite{novello-prd-2012} only that part of the curve $\{\rho_\textsc{b}=\rho_\textsc{b}(F):\,0\leq F\leq F_0=6.53\}$ is physically meaningful. At $F=2.5$ the energy density $\rho_\textsc{b}$ is a maximum in both cases [$\rho^\text{max}_\textsc{b}=1.27$ for (\ref{rho-vs-f}) and $0.27$ for (\ref{l-model'})].}\label{fig1}
\end{figure*}


\section{Extended Born-Infeld theory}

Following \cite{novello-prd-2012}, in order to simplify the analysis, here we shall consider a flat FRW universe filled with a ''magnetic fluid'', i. e., the electric component $E$ will be assumed vanishing or, in other words, only the average of the magnetic part $B$ is different from zero. This choice leads to the so called magnetic universe which is characterized by the following barotropic parameters:\footnote{This situation turns out to be relevant in cosmology as long as the averaged electric field $E$ is screened by the charged primordial plasma, while the magnetic field lines are frozen \cite{lemoine}.}

\bea \rho_\textsc{b}=-L,\;p_\textsc{b}=L-\frac{4}{3}\,L_F\,F,\;F=2B^2.\label{rho-p-b}\eea

Here we shall focus in the study of a cosmological model proposed in \cite{novello-prd-2012}, which is based upon the following modified EM Born-Infeld Lagrangian density \cite{new-bib}:

\bea L=-\gamma^2\,W^{1/2},\;W:=1+\frac{F}{2\gamma^2}-\alpha^2 F^2,\label{l-model}\eea where $\gamma$ and $\alpha$ are free parameters. As seen from Eq. (\ref{rho-p-b}), in a magnetic universe the energy density associated with the EM field 

\bea \rho_\textsc{b}=-L=\gamma^2 \sqrt{1+\frac{F}{2\gamma^2}-\alpha^2 F^2}.\label{rho-vs-f}\eea From this expression it follows that $\rho_\textsc{b}$ vanishes at\footnote{Notice that in Ref. \cite{novello-prd-2012} the values $F_0^\pm$ are incorrectly associated with extrema of the EM field.} 

\bea F=F_0^\pm=\frac{1\pm\sqrt{1+16\alpha^2\gamma^4}}{4\alpha^2\gamma^2},\label{f-max}\eea i. e., the field $F$ is bounded both from above and from below. The value of the field $F=F_0^-$ is unphysical since $F_0^-<0$ is a negative quantity and, for a magnetic universe [$F=2B^2$], only non negative values $$0\leq F\leq F_0\equiv F_0^+,$$ are to be considered (see FIG. \ref{fig1}). The energy density of the magnetic field is a maximum at 

\bea F=F_*=\frac{1}{4\alpha^2\gamma^2}\;\Rightarrow\;\rho_\textsc{b}=\rho^\text{max}_\textsc{b}=\frac{\sqrt{1+16\alpha^2\gamma^4}}{4\alpha}.\label{max}\eea 

A distinctive feature of the model of Ref. \cite{novello-prd-2012} is that, even at vanishing magnetic field $F=0$, the energy density $\rho_\textsc{b}=\gamma^2$ is non null. So that, one wonders, where does the energy density of the EM field come from? The answer is that the present theory does not have the standard (linear) Maxwell electrodynamics limit $L=-F/4$ at $\gamma^2\rightarrow\infty$ ($\alpha=0$). Actually, at this limit one gets from Eq. (\ref{l-model}), $L\approx-\gamma^2-F/4$. But, as discussed in Sec. VII of \cite{novello-prd-2012}, this feature is at the heart of the present model. The question then is, what is the meaning of an EM theory which at weak field does not have the Maxwell limit?

Let us to modify the Born-Infeld Lagrangian (\ref{l-b-i}) just by adding the term quadratic in $F$ within the square root, but without removing the Lagrangian term $\propto \gamma^2$ (see Eq. (31) of Ref. \cite{novello-prd-2012} with $W$ given by Eq. (\ref{l-model})):

\bea &&L=-\gamma^2\left(W^{1/2}-1\right),\nonumber\\
&&\rho_\textsc{b}=\gamma^2\left(\sqrt{1+\frac{F}{2\gamma^2}-\alpha^2 F^2}-1\right).\label{l-model'}\eea Notice that this $\rho_\textsc{b}$ vanishes at vanishing field value $F=0$ ($B=0$) as it should be to recover the correct weak field linear behavior. Contrary to what is stated in \cite{novello-prd-2012}, in this case, the energy density of the magnetic field is non negative definite [$\rho_\textsc{b}\geq 0$] in the finite interval (see the FIG. \ref{fig1}) $$0\leq F\leq\frac{1}{2\alpha^2\gamma^2}\,,$$ where the upper bound on $F$ is obtained, precisely, by requiring the energy density $\rho_\textsc{b}$ to be non negative. Hence, the theory (\ref{l-model'}) meets the requirement which inspired the model of Ref. \cite{novello-prd-2012}: boundedness of the field $F=2B^2$ and of the associated energy density. 

The energy density $\rho_\textsc{b}$ in (\ref{l-model'}) is a maximum at the same field value $F_*$ as $\rho_\textsc{b}$ in Eq. (\ref{rho-vs-f}):

\bea F_*=\frac{1}{4\alpha^2\gamma^2}\;\Rightarrow\;\rho^\text{max}_\textsc{b}=\frac{\sqrt{1+16\alpha^2\gamma^4}-4\alpha\gamma^2}{4\alpha}.\label{max'}\eea Nevertheless, at the maximum, the energy density of the magnetic field is smaller in this case than in the theory of \cite{novello-prd-2012} (compare with Eq. (\ref{max})). 

From this simple analysis it follows that, the main argument stated in Ref. \cite{novello-prd-2012} against the model (\ref{l-model'}) related with the non positivity of $\rho_\textsc{b}$ in (\ref{l-model'}), is not a valid one and that the conclusion about the inevitability of the asymptotic vacuum regime in a magnetic universe is wrong.

\section{Cosmological Evolution}

Assuming plain magnetic universe -- no background matter fluid other than the EM field -- the cosmological equations can be written in the following form:

\bea &&3H^2=\rho_\textsc{b},\;2\dot H=-(\rho_\textsc{b}+p_\textsc{b})=\frac{4}{3}L_F F,\nonumber\\
&&\dot\rho_\textsc{b}+3H(\rho_\textsc{b}+p_\textsc{b})=0,\label{feqs}\eea where $H={\dot a}/a$ is the Hubble parameter, and 

\bea L_F=-\frac{1-4\alpha^2\gamma^2 F}{4\sqrt{W}},\label{l-f}\eea for both models (\ref{l-model}) and (\ref{l-model'}). The continuity equation in (\ref{feqs}) can also be written as 

\bea \dot F+4HF=0\;\Rightarrow\;\dot B=-2HB.\label{eq-1}\eea These equations can be easily integrated

\bea B(a)=B_0 a^{-2},\;F(a)=2B_0^2 a^{-4},\label{b-vs-a}\eea where $B_0$ is an integration constant.

Worth noting that, since 

\bea -2\dot H=\rho_\textsc{b}+p_\textsc{b}=-\frac{4}{3}L_F F,\label{raycha-eq}\eea whenever $L_F F=0$, the magnetic fluid behaves as a cosmological constant and drives the de Sitter evolution of the universe: 

\bea \dot H=0\;\Rightarrow\;F=0\;\left|\right|\;F=F_*=\frac{1}{4\alpha^2\gamma^2}\,.\label{desitter-cond}\eea Hence, at both field values [$F=0$ and $F=F_*$] $H=H_0$. For the model of \cite{novello-prd-2012} at $F=0$, since according to (\ref{rho-vs-f}) $\rho_\textsc{b}(0)=\gamma^2$, one has $H_0=\pm\gamma/\sqrt 3$. Meanwhile, for the model (\ref{l-model'}), at $F=0$, since $\rho_\textsc{b}=0$, then $H_0=0$. In consequence, while for the former theory at vanishing field value one has de Sitter cosmological evolution, for the latter model one has a static universe instead. In both cases [theory (\ref{l-model}) and theory (\ref{l-model'})], at $F=F_*$ where the energy density is a maximum $\rho_\textsc{b}^\text{max}$, the universe is in a stage of de Sitter expansion $$a(t)\propto e^{H_0 t},\;H_0=\sqrt{\rho_\textsc{b}^\text{max}/3}.$$


\begin{figure*}[t]\begin{center}
\includegraphics[width=5.5cm,height=5cm]{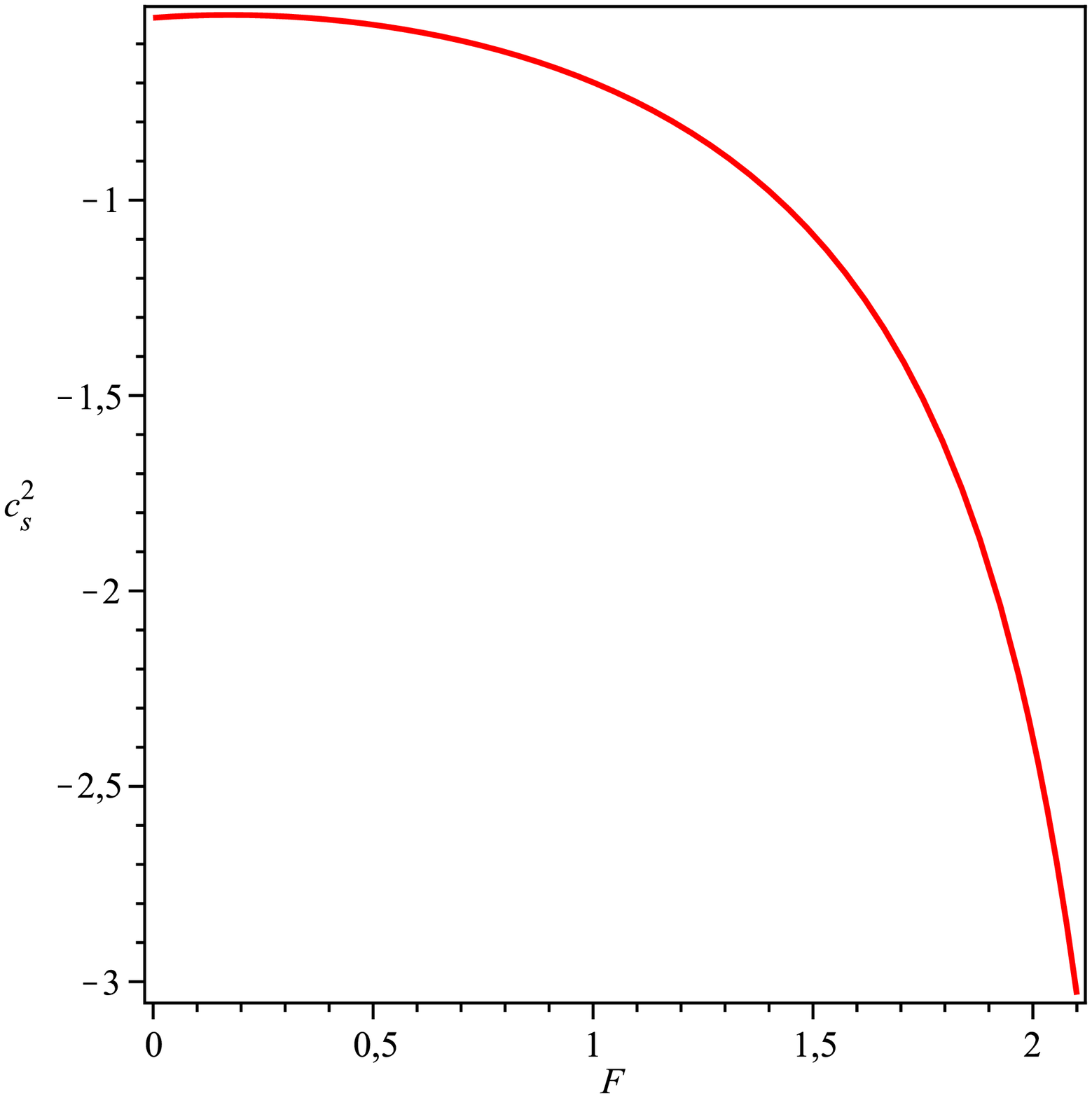}
\includegraphics[width=5.5cm,height=5cm]{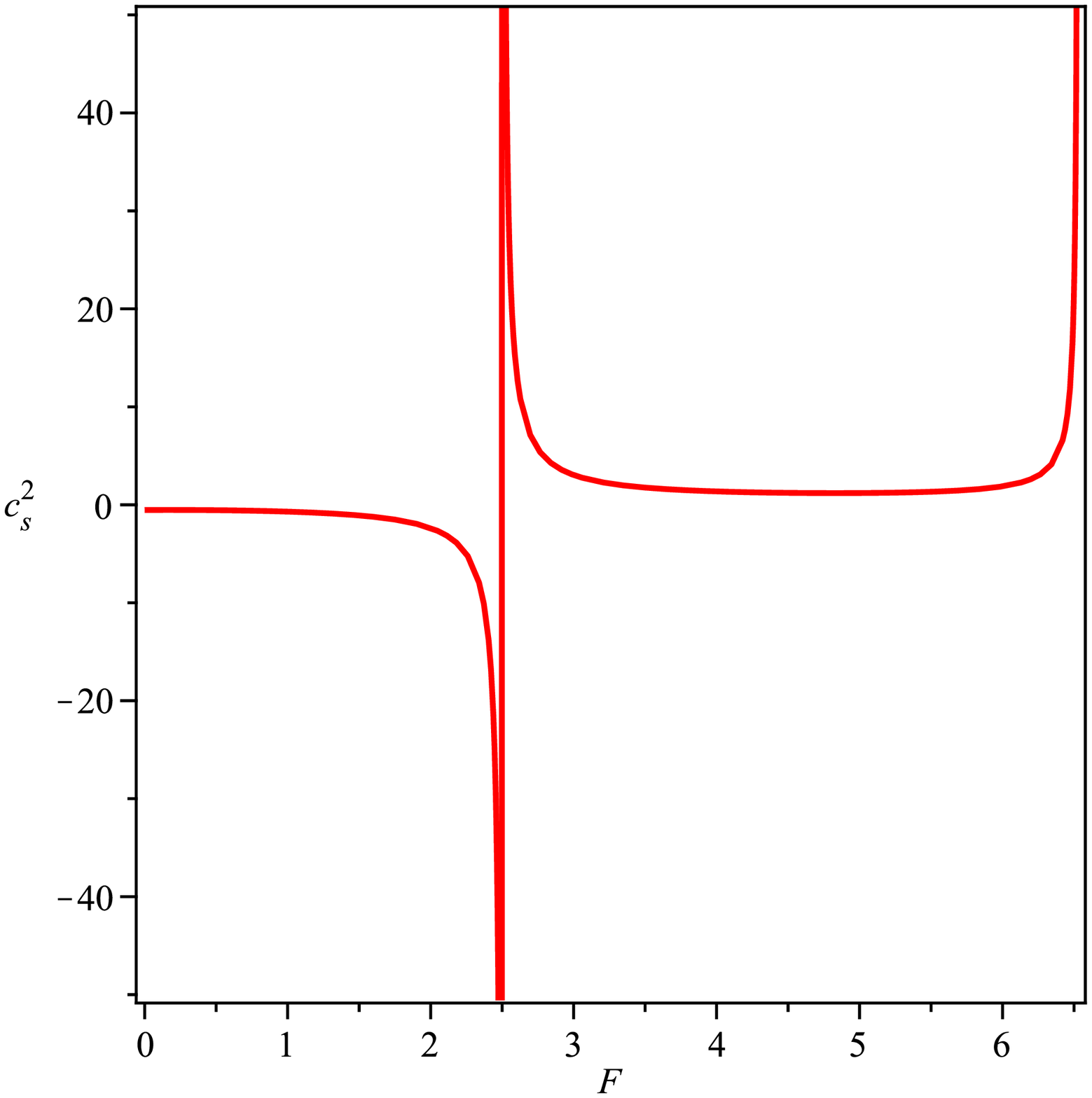}
\includegraphics[width=5.5cm,height=5cm]{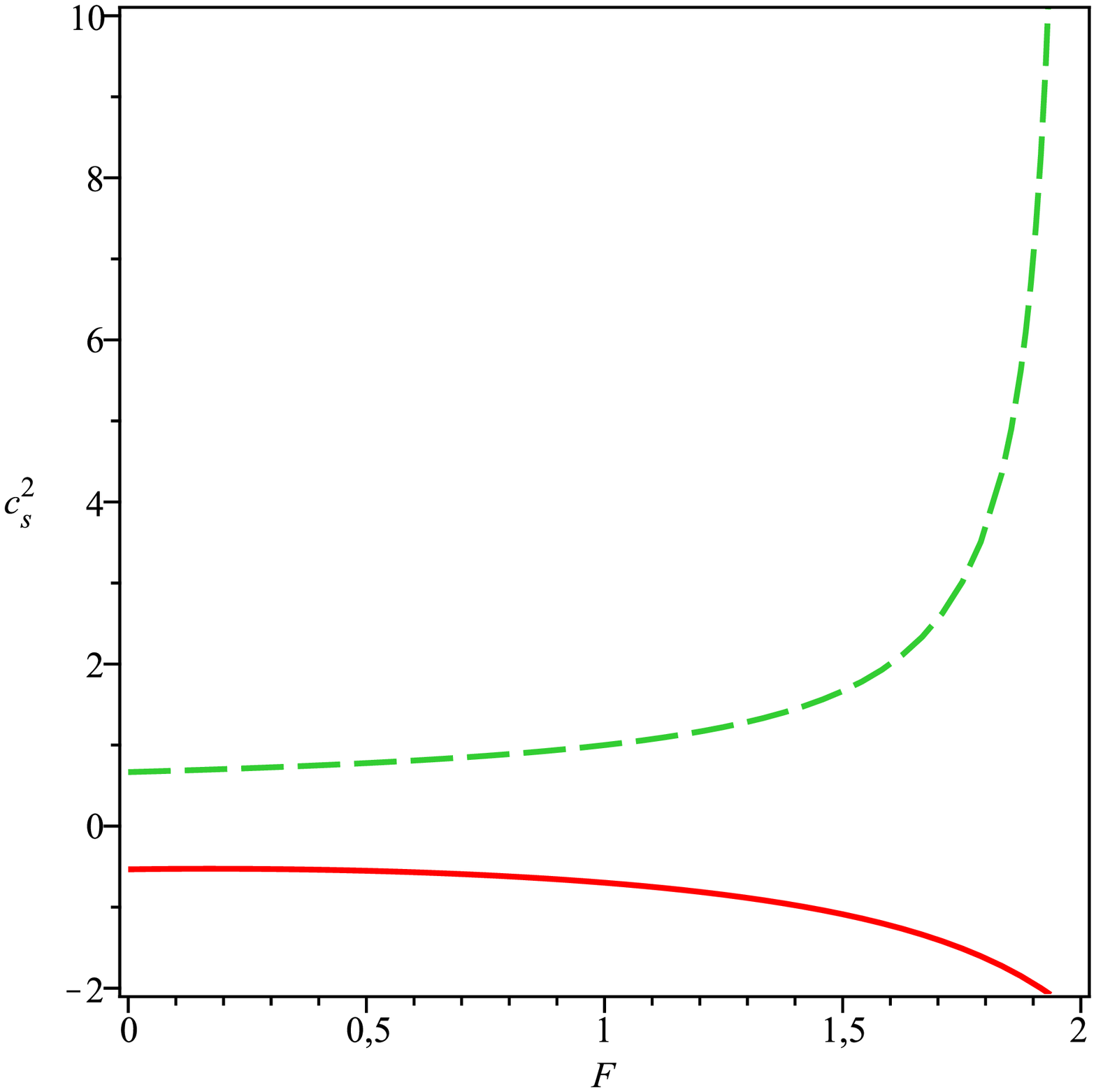}
\end{center}\vspace{0.3in}
\caption{Plot of the speed of sound (squared) $c_s^2$ vs $F$ for arbitrarily chosen $\alpha^2=0.1$ and $\gamma^2=1$. The left-hand and center figures depict $c_s^2=c_s^2(F)$ for the models (\ref{l-model}) and (\ref{l-model'}) -- these coincide in this case -- while the right-hand figure shows $c_s^2(F)$ for the mentioned models (solid curve) and for the model (\ref{l-new}) (dashed curve). It is seen that for the models (\ref{l-model}) and (\ref{l-model'}) the speed of sound is a negative quantity in the field interval $0\leq F<F_*=2.5$. Behind the point $F=F_*$ -- where there is a vertical asymptote and the speed of sound is undefined -- $c_s^2$ for these models is a positive quantity. For the theory of Ref. \cite{novello-prd-2012} $c_s^2$ blows up also at $F=F_0=6.53$ which is associated with a sudden curvature singularity. Meanwhile, for the model (\ref{l-new}) the speed of sound squared is always positive (dashed curve) and blows up at $F=2\gamma^2=2$.}\label{fig3}
\end{figure*}


\section{The bounce}

At the bounce, since the scale factor is a minimum while $H$ changes sign (contraction turns into expansion), then $H=0$. In general, the sufficient conditions for a bounce are \cite{phys-rep-2008}:

\bea \dot a=0,\;\ddot a\geq 0\;\Rightarrow\;H=0,\;\dot H\geq 0,\label{bounce-conditions}\eea where the above quantities are evaluated at the bounce.

For the theory of reference \cite{novello-prd-2012} -- equations (\ref{l-model}), (\ref{rho-vs-f}) -- the condition $H=0$ can not be reconciled with a non-vanishing minimum of the energy density $\rho_\textsc{b}$ in respect to $t$ -- the cosmic time,\footnote{We have to be careful since a given quantity, say $\rho_\textsc{b}=\rho_\textsc{b}(F(t))$, can have extrema in respect to the variable $F$, which might not coincide with extrema of the same quantity in respect to the implicit variable $t$.} as it has been incorrectly stated in \cite{novello-prd-2012} (see section VI.D of the mentioned reference and, in particular, the FIG. 4 where the time evolution of the energy density is shown, and a non-vanishing minimum of $\rho_\textsc{b}$ is associated with the bounce). Actually, at a minimum of $\rho_\textsc{b}$ in respect to $t$, $\dot\rho_\textsc{b}=0$, which means that $$H(\rho_\textsc{b}+p_\textsc{b})=0.$$ This condition is fulfilled even if $\rho_\textsc{b}+p_\textsc{b}\neq 0$, since at the bounce $H=0$. However, according to the Friedmann equation in (\ref{feqs}), $H=0\;\Rightarrow\;\rho_\textsc{b}=0$. Therefor a non-vanishing $\rho_\textsc{b}\neq 0$ at the bounce is not compatible with the Friedmann constraint. This means that in the theory of Ref. \cite{novello-prd-2012} the bounce, if any, can be attained only at vanishing $\rho_\textsc{b}$, i. e., at the upper bound of the $F$-field $F=F_0$, where $W$ vanishes (see equations (\ref{l-model}), (\ref{f-max})). Notice, in passing, that at this point $\dot H\propto L_F F\propto W^{-1/2}$ -- see Eq. (\ref{l-f}) -- blows up, signaling a curvature singularity. Since at $F_0$ $\Rightarrow$ $\rho_\textsc{b}=0$, from $\rho_\textsc{b}+p_\textsc{b}\propto -L_F F$ -- see equation (\ref{feqs}) -- it follows that it is the parametric pressure $p_\textsc{b}$ which blows up at $F=F_0$. This, in addition to the fact that at $F_0$ the scale factor is a finite quantity, means that at this field value a sudden curvature singularity arises. Hence in the theory (\ref{l-model}) exposed in \cite{novello-prd-2012} the bounce, if it exists at all, is to be associated with a curvature singularity, which is not better than the initial cosmological singularity in standard general relativity.

For the theory (\ref{l-model'}) the condition $H=0$ for a bounce occurs at vanishing field $F=0$ [$\dot H=0$] where, consistently with the standard linear Maxwell limit, the associated energy density $\rho_\textsc{b}$ vanishes. Regrettably, since $F\propto a^{-4}$, a vanishing $F$ is associated with an infinitely large value of the scale factor, so that the bounce does not actually occur. In this case the value $F_0$ in Eq. (\ref{f-max}) is never attained. Recall that the upper bound on $F$ is $$\frac{1}{2\alpha^2\gamma}<F_0=\frac{1+\sqrt{1+16\alpha^2\gamma^4}}{4\alpha^2\gamma^2},$$ which means that $\dot H$ (also $H^2$) is always a finite quantity, i. e., the resulting cosmological evolution is regular -- no curvature singularity at all -- unlike for the theory of \cite{novello-prd-2012}.

\section{Discussion and Conclusion}\label{discussion}

In the reference \cite{novello-prd-2012} a generalized Born-Infeld EM theory was proposed which, for a magnetic universe -- according to the authors -- inevitably leads to a cosmological history which interpolates between asymptotic de Sitter states. The main motivation to modify the Born-Infeld Lagrangian (\ref{l-b-i}) in \cite{novello-prd-2012} was to have a regular magnetic universe driven by a bounded field $F=2B^2$. Unfortunately, the resulting theory (\ref{l-model}) does not have the standard weak field (linear) Maxwell limit. To worsen things, as shown in the former section, a sudden curvature singularity -- not better than the big bang singularity of general relativity -- arises at the upper bound of the field $F=F_0$, which is the only field value which could be associated with a bounce (if any) in this theory.

Perhaps a simpler modification of the Born-Infeld theory can be given by the following Lagrangian\footnote{Notice the subtle sign differences with the original Born-Infeld Lagrangian (\ref{l-b-i}).}

\bea L=\gamma^2\left(\sqrt{1-\frac{F}{2\gamma^2}}-1\right).\label{l-new}\eea This theory has the correct linear Maxwell limit at weak field (formal limit $\gamma^2\rightarrow\infty$): $L=-F/4$ and, for a magnetic universe of interest in this comment, allows for an upper bound on the field $F=2\gamma^2$ [$B=\gamma$], which was the motivation of \cite{novello-prd-2012}. In the present case integration of Eq. (\ref{eq-1}) yields to expressing the cosmic time as a monotonic function of the field $F=2B^2$:

\bea &&t(F)-t_0=\frac{\sqrt{3/2}}{2\gamma}\left[\text{arc}\tanh\left(\sqrt\frac{1-\sqrt{1-F/2\gamma^2}}{2}\right)\right.\nonumber\\
&&\left.\;\;\;\;\;\;\;\;\;\;\;\;\;\;\;\;\;\;\;\;\;\;\;\;\;\;\;\;\;\;\;\;\;\;\;\;\;\;+\sqrt\frac{2}{1-\sqrt{1-F/2\gamma^2}}\right],\label{sol-new}\eea where $t_0$ is an integration constant and, to get $F=F(t)$ one has to invert (\ref{sol-new}). The starting point of the cosmic evolution is associated with the upper bound of the field $F=2\gamma^2$ [$B=\gamma$], while at the future infinity $t\rightarrow +\infty$ the field vanishes $F\rightarrow 0$. Regrettably, since for the model (\ref{l-new}) $$p_\textsc{b}=-\rho_\textsc{b}+\frac{F}{3\sqrt{1-F/2\gamma^2}},$$ the start of the cosmological expansion is associated with a sudden singularity: at $F=2\gamma^2$, $H=\gamma/\sqrt 3$ - finite, while $p_\textsc{b}\rightarrow\infty$ blows up (as we shall see below at this field value the speed of sound squared also blows up).

Another quantity of cosmological importance is the speed of sound squared which, for the cases of interest in this comment, can be written as

\bea c_s^2:=\frac{dp_\textsc{b}}{d\rho_\textsc{b}}=\frac{dp_\textsc{b}/dF}{d\rho_\textsc{b}/dF}=\frac{1}{3}+\frac{4L_{FF}}{3L_F}\,F.\label{speed-sound}\eea 

If consider small perturbations of the background energy density $\rho_\textsc{b}(t,{\bf x})=\rho_\textsc{b}(t)+\delta\rho_\textsc{b}(t,{\bf x})$, the conservation of energy-momentum $T^{\mu\nu}_{\;\;;\nu}=0$, leads to the wave equation \cite{peebles-ratra}: $\delta\ddot\rho=c_s^2\nabla^2\delta\rho$, which solution for positive $c_s^2>0$ is $\delta\rho_\textsc{b}=\delta\rho_\textsc{b0}\exp(-i\omega t+i{\bf k}\cdot{\bf x})$ while, for negative $c_s^2<0$, it is $\delta\rho_\textsc{b}=\delta\rho_\textsc{b0}\exp(\omega t+i{\bf k}\cdot{\bf x})$. In the latter case the energy density perturbation uncontrollably grows resulting in an instability of the cosmological model. For the models (\ref{l-model}) and (\ref{l-model'}) the speed of sound is the same $$c_s^2=\frac{1}{3\gamma^2}\left[\gamma^2-\frac{1+16\alpha^2\gamma^4}{(1-4\alpha^2\gamma^2F)W}\right],$$ where $W$ is defined in (\ref{l-model}), while for the model (\ref{l-new}) $$c_s^2=\frac{1}{3\gamma^2}\left(\gamma^2+\frac{1}{1-F/2\gamma^2}\right).$$ 

A plot of $c_s^2$ vs $F$ is shown in the FIG. \ref{fig3} for the theories (\ref{l-model}), (\ref{l-model'}) -- solid curves -- and (\ref{l-new}) -- dashed curve -- for arbitrarily chosen $\alpha^2=0.1$ and $\gamma^2=1$. It is seen that for theories (\ref{l-model}) and (\ref{l-model'}) the speed of sound squared is negative in the field interval $0\leq F\leq F_*$ and that, at $F=F_*$ where the magnetic background energy density is a maximum, $c_s^2$ is undefined (there is a vertical asymptote). This signals a fundamental instability against small perturbations of the background energy density in a magnetic universe described by both theories in this field interval. For the theory of \cite{novello-prd-2012} the speed of sound is also undefined at $F=F_0$ where, as shown in the former section, a sudden curvature singularity arises. For the most simple theory (\ref{l-new}) the speed of sound squared $c_s^2$ is always a non negative quantity and blows up at $F=2\gamma^2$, where also a sudden curvature singularity occurs.

Our main conclusion is that none of the modifications of the Born-Infeld theory where $1+F/2\gamma^2$ is replaced by $1+F/2\gamma^2-\alpha^2 F^2$ -- theories given by (\ref{l-model}) and (\ref{l-model'}) -- can be an adequate cosmological model since, thanks to the fact that the speed of sound squared can be negative, the resulting theory is unstable against small perturbations of the background energy density. The most simple such modification of the Born-Infeld theory which (i) has the correct weak field Maxwell limit, (ii) for a magnetic universe the field $F$ is bounded, and (iii) is free of curvature singularities, is the one given by the Lagrangian (\ref{l-model'}), which was incorrectly ruled out in \cite{novello-prd-2012} but which, nevertheless, is to be ruled out due to the mentioned instability. In this theory the fate of the cosmic evolution is a static universe and, besides, there is no place for the bounce. The alternative theory (\ref{l-new}) is free of the instability against small perturbations of the background energy density although a sudden curvature singularity is also unavoidable in this theory. 

Summarizing: the main conclusion of \cite{novello-prd-2012} about the inevitability of the asymptotic vacuum regime in a magnetic universe is wrong. Besides, the theory of Ref. \cite{novello-prd-2012}; (i) does not have the Maxwell limit, (ii) in a cosmological setting a sudden curvature singularity is inevitable in it, and (iii) it is unstable against small perturbations of the background magnetic energy density, so that it should be ruled out.

The authors thank SNI of Mexico for support. The work of R G-S was partly supported by SIP20131811, COFAA-IPN, and EDI-IPN grants. I Q thanks "Programa PRO-SNI, Universidad de Guadalajara" for support under grant No 146912.

\end{document}